\documentclass[11pt,a4paper]{article}
\usepackage[utf8]{inputenc}
\usepackage{amsmath}
\usepackage{amsfonts}
\usepackage{amssymb}
\usepackage{graphicx}
\usepackage{hyperref}
\usepackage{booktabs}
\usepackage{geometry}
\usepackage[ruled,vlined,linesnumbered]{algorithm2e}
\usepackage{float}
\usepackage{enumitem}
\usepackage{tabularx}
\usepackage{natbib}
\title{Contextual Memory Virtualisation: DAG-Based State Management and Structurally Lossless Trimming for LLM Agents}

\author{
    Cosmo Santoni$^{1,2}$ \\
    $^1$Imperial College London \\
    $^2$Ingression Labs \\
    \texttt{cosmo.santoni@imperial.ac.uk}, \texttt{cosmo@ingressionlabs.com}
}
\date{February 2026}

\begin{document}

\maketitle

\begin{abstract}
As large language models engage in extended reasoning tasks, they accumulate significant state -- architectural mappings, trade-off decisions, codebase conventions -- within the context window. This understanding is lost when sessions reach context limits and undergo lossy compaction. We propose contextual memory virtualisation (CMV), a system that treats accumulated LLM understanding as version-controlled state. Borrowing from operating system virtual memory, CMV models session history as a Directed Acyclic Graph (DAG) with formally defined snapshot, branch, and trim primitives that enable context reuse across independent parallel sessions. We introduce a three-pass structurally lossless trimming algorithm that preserves every user message and assistant response verbatim while reducing token counts by a mean of 20\% and up to 86\% for sessions with significant overhead by stripping mechanical bloat such as raw tool outputs, base64 images, and metadata. A single-user case-study evaluation across 76 real-world coding sessions demonstrates that trimming remains economically viable under prompt caching, with the strongest gains in mixed tool-use sessions, which average 39\% reduction and reach break-even within 10 turns. A reference implementation is available at \url{https://github.com/IngressionLabs/contextual-memory-virtualisation}.
\end{abstract}

\section{Introduction}

\begin{figure}[t]
\centering
\includegraphics[width=0.48\textwidth]{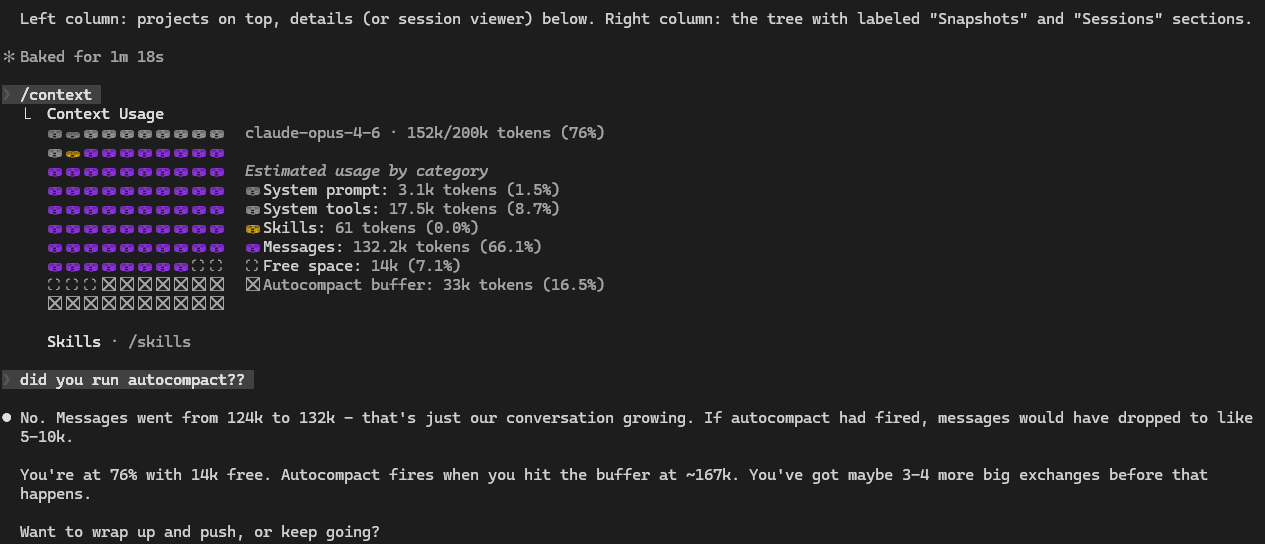}
\hfill
\includegraphics[width=0.48\textwidth]{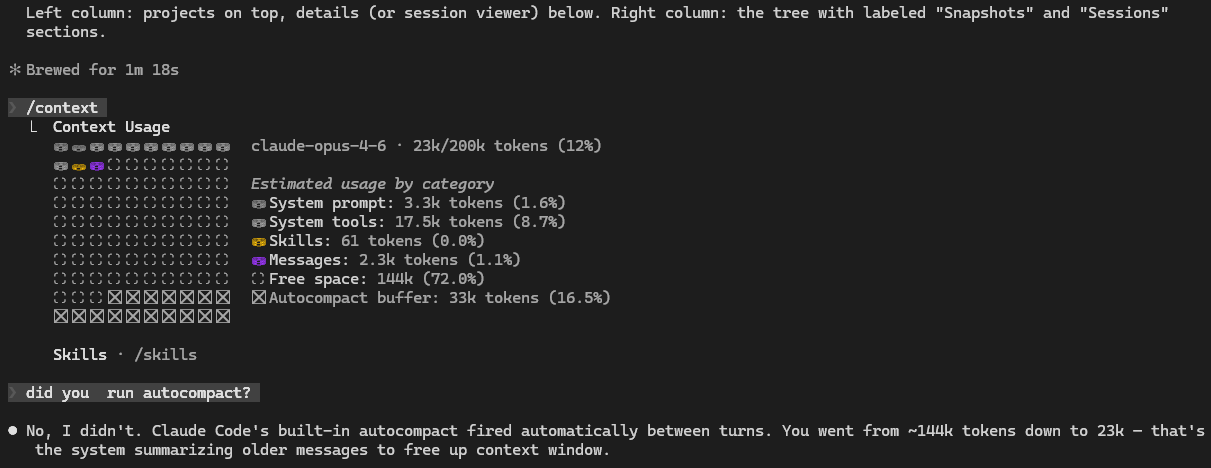}
\caption{Context window before (left, 132k message tokens, 76\% capacity) and after (right, 2.3k message tokens, 12\% capacity) native autocompaction (e.g., Claude Code). Autocompaction summarises 98\% of accumulated session state into a brief summary to reclaim window space.}
\label{fig:compaction}
\end{figure}

Extended work sessions with LLM coding agents build cumulative state within the context window. Architecture gets mapped, trade-offs get weighed, decisions are recorded, conventions are learned. After 30 minutes of deep work, the model holds a mental model of an entire codebase, built at significant computational cost in both time and tokens. When the context window fills, native compaction (e.g., \texttt{/compact} in Claude Code \citep{anthropic2024claude}) summarises this state into a few sentences. In observed sessions, autocompaction reduced 132k tokens of accumulated message state to 2.3k---a 98\% reduction (Figure~\ref{fig:compaction})---discarding the nuanced understanding that took an entire session to build. Each new session starts from scratch.

This is a fundamental inefficiency. The cost of building context is paid repeatedly, and the resulting understanding is never preserved in a reusable form. Existing approaches address fragments of this problem. Retrieval-Augmented Generation (RAG) \citep{lewis2020rag} augments prompts with retrieved documents but does not preserve conversational state. MemGPT \citep{packer2023memgpt} applies OS-inspired paging to swap context segments in and out of the window, but is limited to a single session and relies on the model to manage its own memory. Memory plugins persist summary facts across sessions but lose the full conversational nuance. Native session utilities (\texttt{/rewind}, \texttt{--fork}) provide within-session undo and one-off copies but lack named states, lineage tracking, or context cleanup.

A separate line of work addresses context window pressure through prompt compression. LongLLMLingua \citep{jiang2023longllmlingua} accelerates inference by compressing long prompts via perplexity-guided token pruning. Chevalier et~al.\ \citep{chevalier2023adapting} train models to produce compressed soft tokens from long contexts. RECOMP \citep{xu2024recomp} selectively compresses retrieved documents before augmentation. Ge et~al.\ \citep{ge2024context} propose an in-context autoencoder that learns to compress and reconstruct context segments. These approaches modify the representation of context at the model or embedding level. Contextual memory virtualisation (CMV) operates at a different layer entirely: it manages the raw conversation log, preserving full fidelity while removing structural bloat. The two approaches are complementary. The attention mechanism underlying modern LLMs \citep{vaswani2017attention} processes all tokens in the context window with equal cost, making window size reduction valuable regardless of the method used.

We frame this solution as \emph{contextual memory virtualisation}. Just as virtual memory in an operating system abstracts away the physical limits of hardware RAM---giving each process the illusion of a vast, contiguous memory space via paging---CMV abstracts away the strict physical token limits of the LLM context window. Instead of forcing the model to live entirely within its "RAM" (the current window), CMV allows the user to page saved architectural understanding in and out of active context as needed, effectively decoupling the cost of building context from the cost of executing a task.

CMV comprises three contributions. At its core is a DAG-based state model that formalises context snapshots as nodes and branches as edges, allowing a single context-building session to act as a persistent root for multiple independent workstreams. We pair this with a three-pass structurally lossless trimming algorithm that safely strips mechanical overhead---such as raw tool outputs and base64 images---while keeping all user and assistant messages intact and handling orphaned tool results to maintain API correctness. Finally, we provide an empirical cost analysis across 76 real-world sessions to demonstrate that this trimming approach remains economically viable even under prompt caching penalties. The reference implementation targets Claude Code, but the DAG model and trimming architecture are agent-agnostic; any system that stores conversation logs and uses tool-call schemas can apply the same approach.

\section{The DAG State Model}

\subsection{Formal Definition}

We model session history as a DAG $G = (V, E)$ where:
\begin{itemize}[nosep]
    \item Each node $v \in V$ is a \emph{snapshot}: an immutable copy of a session's JSONL conversation log at a point in time, annotated with metadata (name, timestamp, source session, estimated tokens, tags).
    \item Each directed edge $(v_i, v_j) \in E$ represents a \emph{branch}: an independent work session forked from snapshot $v_i$ that eventually yields snapshot $v_j$. The forked session receives a copy of $v_i$'s conversation state (optionally trimmed) and a fresh session identifier.
\end{itemize}

A snapshot $v$ may have a \emph{parent snapshot} $\text{parent}(v)$ if the session from which $v$ was captured was itself branched from an earlier snapshot. This induces a lineage chain: $v_0 \to v_1 \to \cdots \to v_k$, where each $v_{i+1}$ inherits the cumulative understanding of all ancestors. This branching structure forms a directed tree (a strict subclass of Directed Acyclic Graphs). We adopt the broader DAG terminology to align with version-control conventions and to accommodate future merge primitives.

\subsection{Core Operations}

Four primitives operate on this graph:

\textbf{Snapshot}($s$) $\to v$: Given a session $s$, copies the JSONL conversation file to immutable storage and creates a new node $v$ with metadata. The original session is never modified.

\textbf{Branch}($v$, trim) $\to s'$: Given a snapshot $v$, creates a new session $s'$ with a fresh UUID. If trim is enabled (the default), the conversation is processed by the trimming algorithm (Section~\ref{sec:trimming}) before being written to the new session. An optional orientation message can be prepended as the first user line to point the model toward a specific task on the new branch.

\textbf{Trim}($s$) $\to s'$: A convenience operation that composes Snapshot and Branch: captures the current session, trims it, and launches a new session in one step.

\textbf{Tree}($G$) $\to$ visualisation: Traverses parent links to reconstruct the full DAG and renders it with ASCII connectors, providing a \texttt{git log --graph} equivalent for conversational context.

\subsection{Practical Implications}

By modeling conversation state as a DAG, CMV introduces a \emph{version-control paradigm for LLM context}---effectively a Git-like workflow for conversational memory. Previously, interacting with an LLM was strictly linear and ephemeral: a single thread that inevitably degrades upon compaction. Under CMV, a user who spends 40 minutes generating 80k tokens of architectural understanding can snapshot that state as a stable root commit. From this root, they can spawn independent, parallel branches for authentication work, API refactoring, or performance tuning, without ever repeating the context-building phase.

\section{Three-Pass Structurally Lossless Trimming}\label{sec:trimming}


The core technical challenge is reducing the token payload without losing the model's synthesised understanding. Inspection of real session data reveals that the majority of context window usage is consumed by \emph{mechanical overhead} (raw file dumps returned as tool results, base64-encoded images, thinking block signatures, file-history metadata) rather than by the conversation itself. The model's synthesis of these inputs (its architectural summaries, design decisions, and explanations) is contained in assistant response blocks, which are typically a small fraction of total tokens.

We introduce a streaming algorithm that strips this mechanical overhead while preserving every user message and assistant response verbatim. If the model needs a file's contents again after trimming, it simply re-reads the file.

\subsection{Algorithm Architecture} 

The trimmer processes JSONL-formatted conversation logs in three sequential passes, as outlined in Algorithm~\ref{alg:trimmer}. Pass~1 and Pass~2 are cheap preparatory scans; Pass~3 performs the actual filtering and writes the output.

\begin{algorithm}[H]
\caption{Three-Pass Structurally Lossless Trimmer}\label{alg:trimmer}
\DontPrintSemicolon
\KwIn{source JSONL path $S$, stub threshold $\tau$ (default 500 chars)}
\KwOut{trimmed JSONL path $D$, metrics $M$}
\BlankLine
\tcp{Pass 1: Compaction Boundary Detection}
$B \gets -1$\;
\ForEach{line $\ell_i$ in $S$}{
    \If{\textnormal{\texttt{String.includes()}} matches compaction markers}{
        parse $\ell_i$\;
        \If{type $\in \{$\texttt{summary}, \texttt{compact\_boundary}$\}$}{
            $B \gets i$\;
        }
    }
}
\BlankLine
\tcp{Pass 2: Pre-Boundary Tool ID Collection}
$\mathcal{O} \gets \emptyset$\;
\ForEach{line $\ell_i$ in $S$ where $i < B$}{
    \ForEach{content block $b$ in $\ell_i$}{
        \If{$b.\text{type} = $ \texttt{tool\_use}}{
            $\mathcal{O} \gets \mathcal{O} \cup \{b.\text{id}\}$\;
        }
    }
}
\BlankLine
\tcp{Pass 3: Stream-Process with Trim Rules}
\ForEach{line $\ell_i$ in $S$}{
    \If{$i < B$}{skip (pre-compaction content)\;}
    \If{type $\in \{$\texttt{file-history}, \texttt{queue-op}$\}$}{skip\;}
    Strip base64 image blocks\;
    Remove thinking blocks (non-portable signatures)\;
    Stub \texttt{tool\_result} content $> \tau$ chars\;
    Stub write-tool inputs $> \tau$ chars (preserve metadata fields)\;
    Strip \texttt{tool\_result} blocks where $\text{id} \in \mathcal{O}$ (orphans)\;
    Remove API usage metadata\;
    Write processed $\ell_i$ to $D$\;
}
\Return{$D, M$}\;
\end{algorithm}

Pass~1 uses \texttt{String.includes()} on raw lines to detect potential compaction boundaries \emph{without} parsing JSON on every line, making the scan near-costless on large files. Only matching lines are parsed. Pass~2 collects tool\_use IDs that will be needed in Pass~3 for orphan detection (Section~\ref{sec:orphan}).

\subsection{Trim Rules and Preservation Guarantees}

The algorithm applies the following rules during Pass~3. Critically, every user message, every assistant response, and every tool \emph{request} (the invocation metadata) is preserved verbatim. Only mechanical \emph{outputs} are reduced:

\begin{itemize}[nosep]
    \item \textbf{Pre-compaction skip:} All lines before the last compaction boundary are discarded (already summarised by native compaction).
    \item \textbf{Metadata removal:} \texttt{file-history-snapshot} and \texttt{queue-operation} entries are discarded.
    \item \textbf{Image stripping:} Base64 image blocks are removed unconditionally.
    \item \textbf{Tool result stubbing:} \texttt{tool\_result} content exceeding $\tau$ characters is replaced with a stub: \texttt{[Trimmed: \textasciitilde N chars]}.
    \item \textbf{Tool input stubbing:} For write-oriented tools, large \texttt{content}, \texttt{old\_string}, and \texttt{new\_string} fields are stubbed. A whitelist of metadata fields (\texttt{file\_path}, \texttt{command}, \texttt{description}, \texttt{path}, \texttt{url}, etc.) is never stubbed, ensuring the model retains knowledge of \emph{which} files were read and \emph{which} commands were run.
    \item \textbf{Thinking block removal:} Thinking blocks require a cryptographic signature that is not portable across sessions and are removed entirely.
    \item \textbf{Orphaned tool result stripping:} Detailed in Section~\ref{sec:orphan}.
\end{itemize}

The stub threshold $\tau$ defaults to 500 characters (minimum 50) and is configurable per-operation.

\subsection{Orphaned Tool Result Handling}
\label{sec:orphan}

LLM tool-use APIs typically enforce a strict schema: every \texttt{tool\_result} block must reference a \texttt{tool\_use} block present in a preceding assistant message. Native compaction often places its boundary between a tool invocation (in an assistant turn) and the corresponding result (in the next user turn). When pre-boundary content is discarded in Pass~3, the tool\_use blocks that lived before the boundary are removed, but their corresponding tool\_result blocks may exist \emph{after} the boundary. Without correction, submitting a session containing these ``orphaned'' results causes an API validation error and the session cannot be resumed.

Pass~2 collects the set $\mathcal{O}$ of all tool\_use IDs from pre-boundary content. During Pass~3, any tool\_result whose \texttt{tool\_use\_id} $\in \mathcal{O}$ is silently discarded. This maintains API correctness without user intervention and was the primary motivation for the three-pass architecture.

\section{Economic Evaluation}

Major LLM APIs implement prompt caching (e.g., \citealt{anthropic2024caching}). If the prompt prefix matches a previously cached prefix, cached tokens are read at a reduced rate rather than reprocessed at the write rate. Trimming necessarily changes the prefix, invalidating the cache and incurring a one-time miss penalty. We evaluate whether per-turn savings from caching a smaller prefix recover that penalty.

\subsection{Cost Model}

For a cache hit rate $h$, the steady-state cost per turn at token count $T$ is:
\begin{equation*}
C(T, h) = \frac{T}{10^6} \left( h \cdot P_{\text{read}} + (1-h) \cdot P_{\text{write}} \right)
\end{equation*}
where $P_{\text{read}}$ and $P_{\text{write}}$ are the per-million-token cache read and write prices respectively. The first turn after a trim incurs a cold-cache penalty at the full write rate:
\begin{equation*}
C_{\text{cold}}(T) = \frac{T}{10^6} \cdot P_{\text{write}}
\end{equation*}

The one-time penalty and per-turn savings are:
\begin{align*}
\Delta_{\text{penalty}} &= C_{\text{cold}}(T_{\text{post}}) - C(T_{\text{pre}}, h) \\
\Delta_{\text{savings}} &= C(T_{\text{pre}}, h) - C(T_{\text{post}}, h)
\end{align*}

Break-even occurs at turn $n^*$, where turn~1 is the initial cold-cache turn:
\begin{equation*}
n^* = \left\lceil \frac{\Delta_{\text{penalty}}}{\Delta_{\text{savings}}} \right\rceil + 1
\end{equation*}

\subsection{Methodology}

We scanned sessions from a single user's Claude Code installation (running Claude Opus 4.6) over a three-month period, excluding internal subagent sessions and sessions with fewer than 10 messages or 5{,}000 tokens, yielding 76 qualifying sessions. Sessions were categorised by bloat profile based on tool result bytes as a proportion of total JSONL bytes: \emph{mixed} ($\geq 15\%$) and \emph{conversational} ($<15\%$). Token counts are derived from byte counts via a chars/4 heuristic plus a fixed system overhead estimate. This overestimates reduction for image-heavy sessions, where the API charges a fixed vision-token cost ($\sim$1{,}600 tokens) independent of base64 encoding size; text-dominated sessions (the majority of the corpus) are unaffected. We assume a steady-state cache hit rate of $h = 0.9$ and report results under Opus~4.6 pricing. Break-even values are capped at 60 turns as a practical planning horizon; sessions with negligible reduction yield arbitrarily large raw break-even values that are not operationally meaningful.

\subsection{Results}

Table~\ref{tab:pricing} shows the pricing model used for the primary analysis. Table~\ref{tab:results} summarises overall trimming results, and Table~\ref{tab:tiers} segments results by session bloat profile.

\begin{table}[H]
\centering
\caption{API pricing per million tokens as of February 2026. Cache write cost is $1.25\times$ base input. Break-even results scale linearly across pricing tiers.}
\label{tab:pricing}
\small
\begin{tabular}{lrrr}
\toprule
\textbf{Model} & \textbf{Base Input} & \textbf{Cache Write} & \textbf{Cache Read} \\
\midrule
Opus 4.6     & \$5.00  & \$6.25  & \$0.50 \\
\bottomrule
\end{tabular}
\end{table}

\begin{table}[H]
\centering
\caption{Trimming results across 76 sessions from a single API-key user (Opus~4.6, $h = 0.9$). Column extrema are independent; the 0\% minimum reduction yields the 60-turn cap, while the 1-turn minimum break-even corresponds to the session with maximum reduction. The negative minimum penalty arises when the trimmed session is small enough that its cold-cache cost is lower than the pre-trim steady-state cost.}
\label{tab:results}
\small
\begin{tabular}{lrrrr}
\toprule
\textbf{Metric} & \textbf{Min} & \textbf{Median} & \textbf{Mean} & \textbf{Max} \\
\midrule
Token reduction (\%)          & 0    & 12   & 20   & 86 \\
Cache miss penalty (\$)       & $-$0.02 & 0.32 & 0.30 & 0.82 \\
Break-even (turns, Opus 4.6)  & 1    & 38   & 35   & 60 \\
\bottomrule
\end{tabular}
\vspace{2pt}
\end{table}

\begin{table}[H]
\centering
\caption{Trimming results segmented by session bloat profile (tool result bytes as \% of total JSONL bytes).}
\label{tab:tiers}
\small
\begin{tabular}{lrrrrr}
\toprule
\textbf{Bloat Profile} & \textbf{Sessions} & \textbf{Mean Red.} & \textbf{Median Red.} & \textbf{Mean Break-even} & \textbf{Mean Context} \\
\midrule
Mixed ($\geq$15\%) & 12 & 39\% & 33\% & 10 turns & 97k \\
Conversational ($<$15\%) & 64 & 17\% & 10\% & 40 turns & 82k \\
\midrule
\textbf{All sessions} & \textbf{76} & \textbf{20\%} & \textbf{12\%} & \textbf{35 turns} & \textbf{84k} \\
\bottomrule
\end{tabular}
\end{table}

\begin{figure}[H]
\centering
\includegraphics[width=0.75\textwidth]{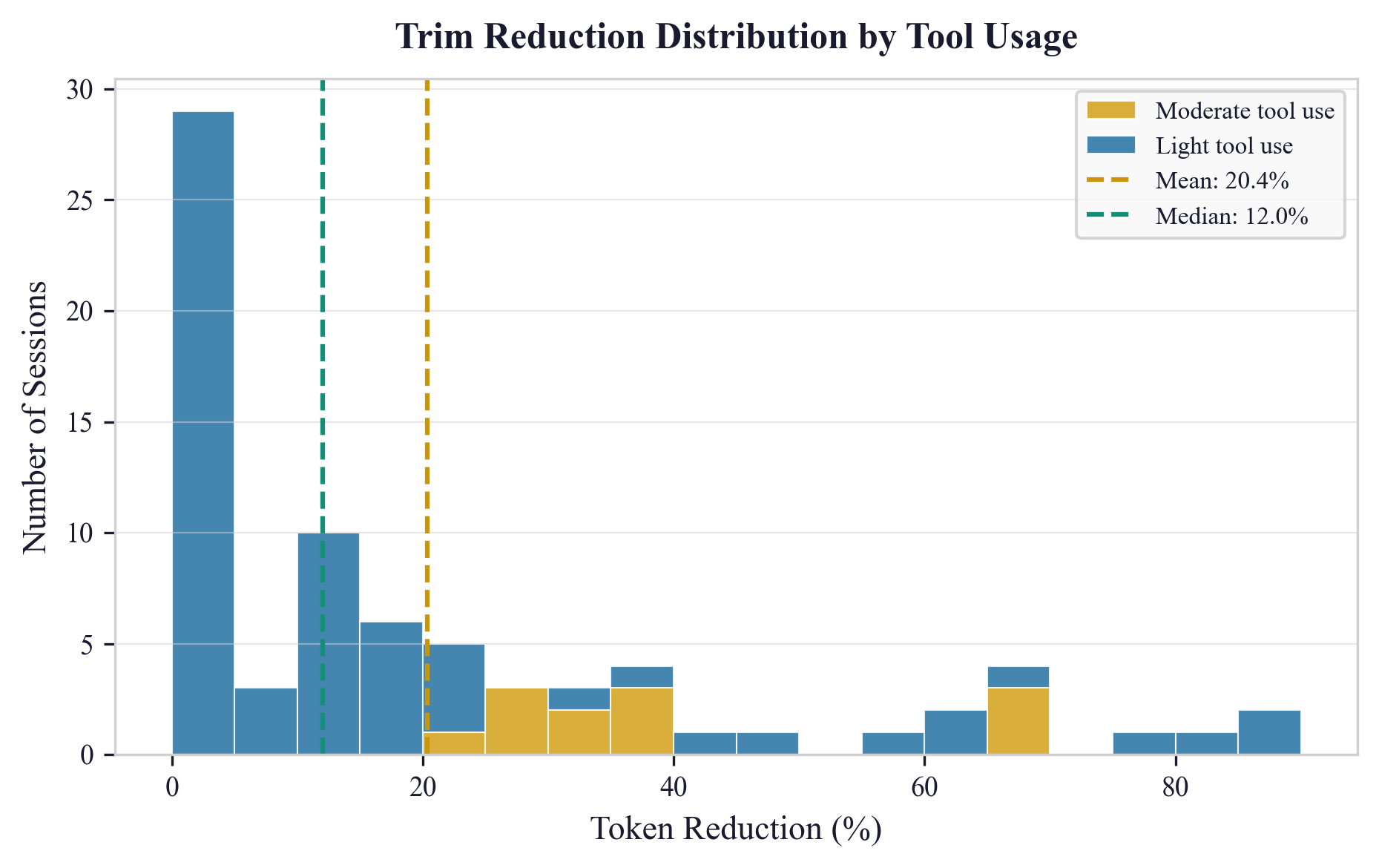}
\caption{Distribution of token reduction across 76 sessions, segmented by bloat profile. The median reduction is 12\%; the mean is pulled higher (20\%) by a tail of sessions with significant trimmable overhead.}
\label{fig:reduction}
\end{figure}

\begin{figure}[H]
\centering
\includegraphics[width=0.75\textwidth]{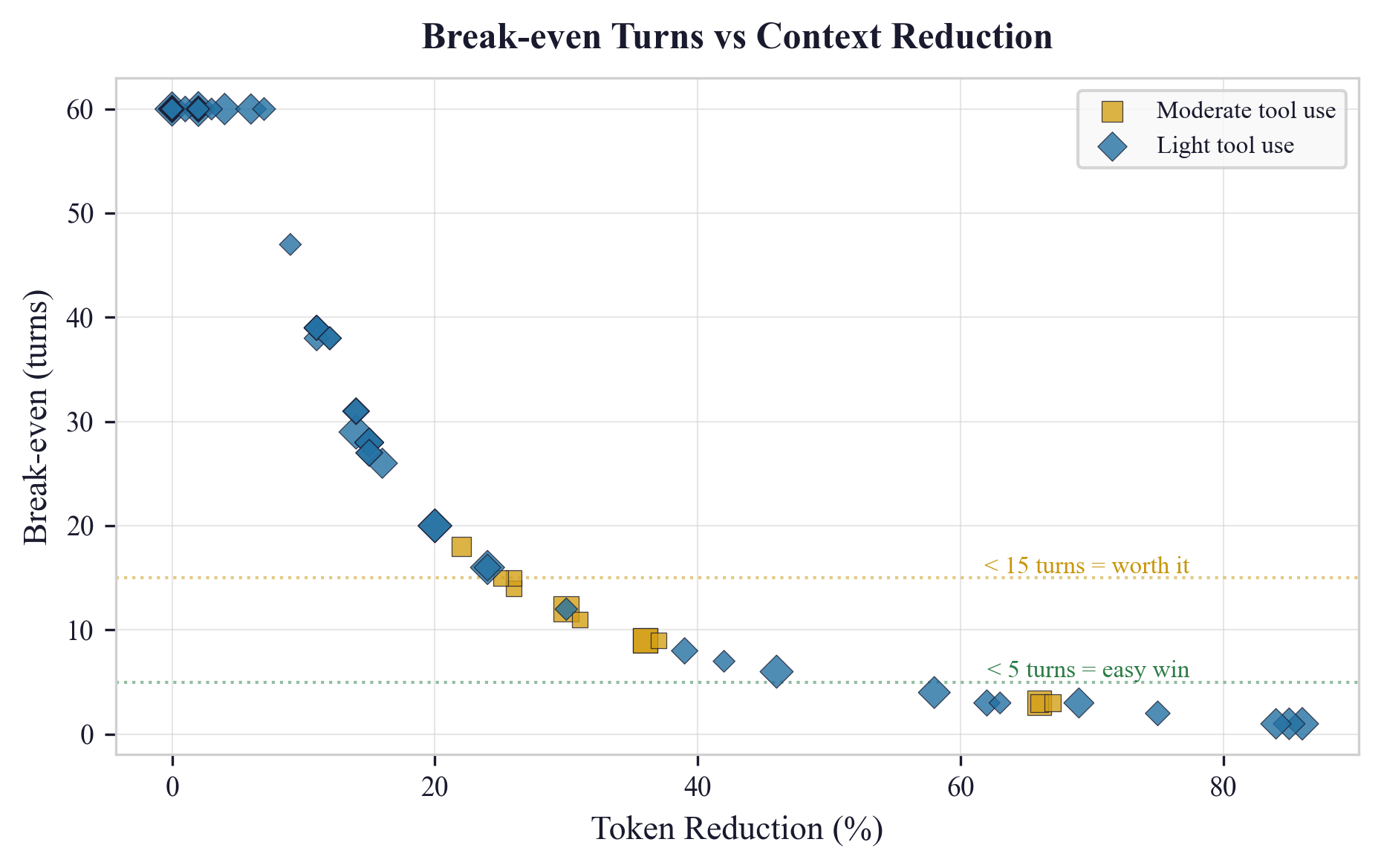}
\caption{Break-even turns vs.\ token reduction. Sessions with $>$30\% reduction reach break-even within 15 turns. Sessions with minimal overhead cluster at the 60-turn cap, correctly indicating trimming is unnecessary.}
\label{fig:breakeven}
\end{figure}

Figure~\ref{fig:reduction} shows the distribution of token reduction across sessions. The majority of sessions are conversational with modest trim gains, but a long tail of sessions with significant trimmable overhead achieves reductions of 40--86\%. Figure~\ref{fig:breakeven} shows the relationship between reduction and break-even: sessions above 30\% reduction reach break-even within 15 turns. The highest-reduction sessions (60--86\%) are driven primarily by pre-compaction history skipping rather than tool result stubbing, indicating two distinct reduction modes: sessions that accumulated large pre-compaction logs benefit from boundary detection, while mixed-profile sessions benefit from tool output and metadata stripping.

\begin{figure}[H]
\centering
\includegraphics[width=0.75\textwidth]{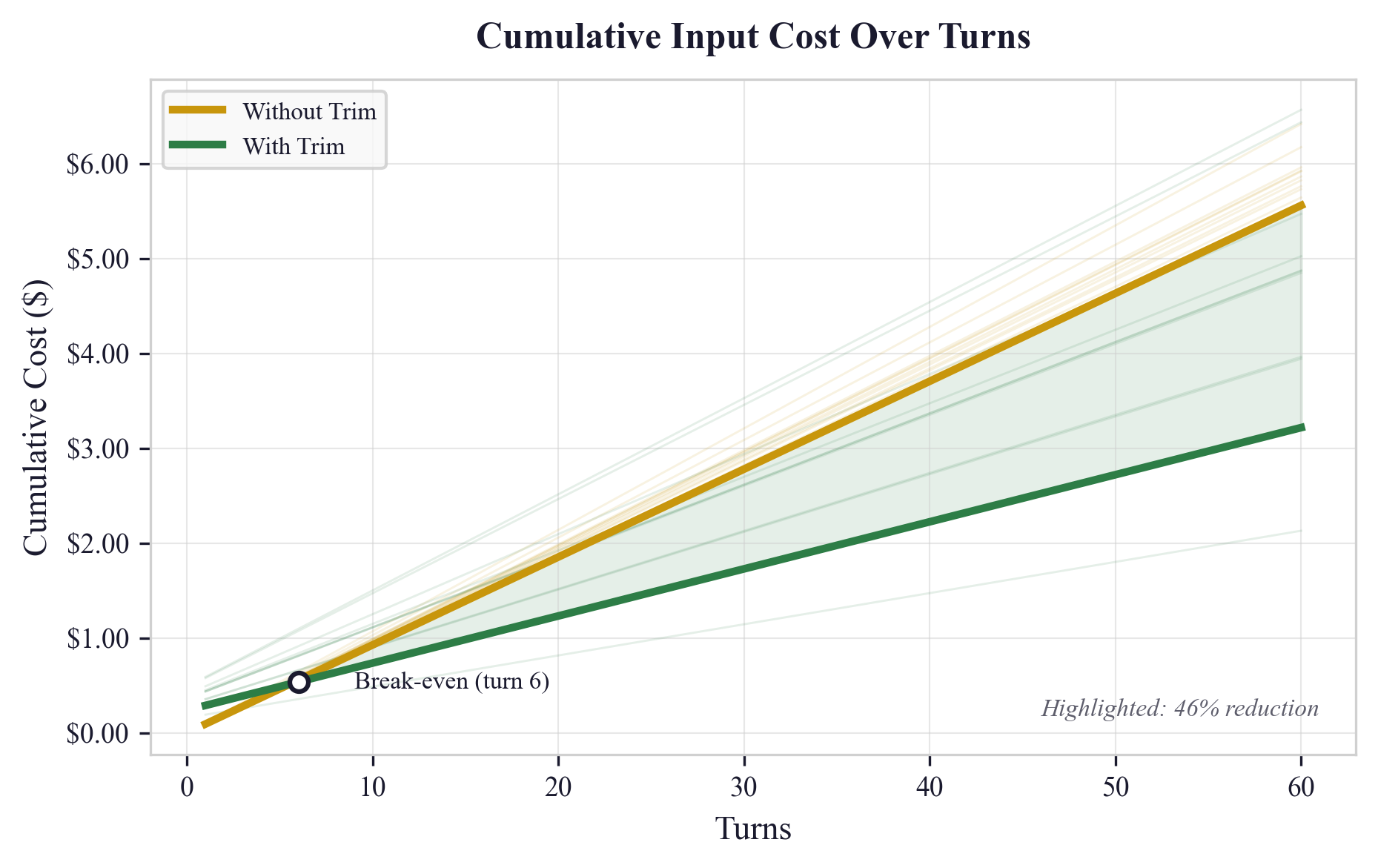}
\caption{Cumulative input cost with and without trimming. The highlighted session (46\% reduction) reaches break-even at turn~6. Faint lines show other sessions; sessions with greater reduction diverge earlier, while sessions with minimal reduction show negligible separation.}
\label{fig:costcurves}
\end{figure}

Figure~\ref{fig:costcurves} illustrates cumulative input cost for a representative session with 46\% reduction. The trimmed session incurs a higher first-turn cost (cold cache), but the lower per-turn rate causes the curves to diverge, with break-even at turn~6. Faint lines show other sessions in the corpus; the spread reflects the range of reduction percentages.

\begin{figure}[H]
\centering
\includegraphics[width=\textwidth]{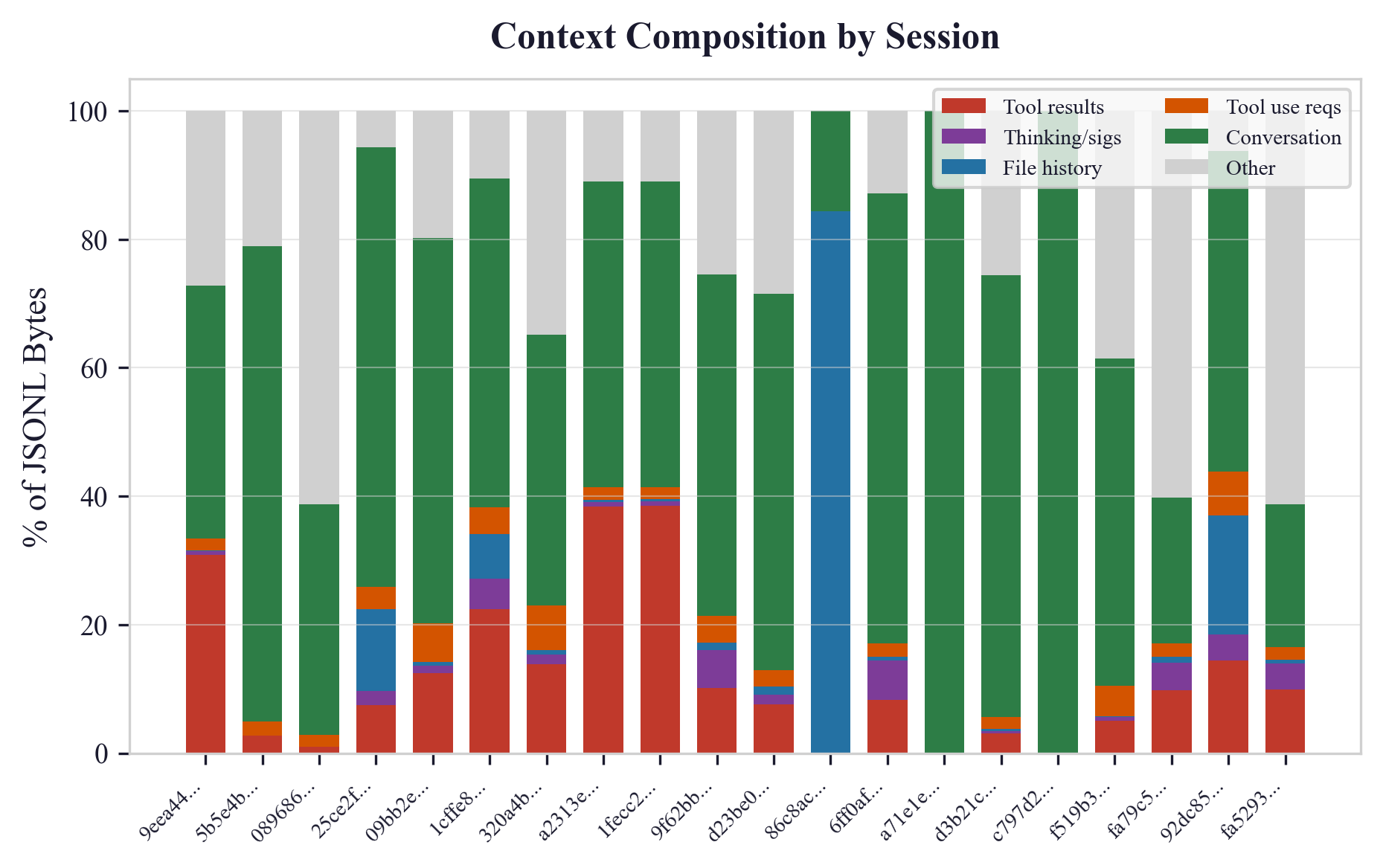}
\caption{Context composition by session. Green represents conversation content (preserved by trimming); red, orange, purple, and blue represent trimmable overhead. Sessions with more overhead see larger reductions.}
\label{fig:composition}
\end{figure}

The composition of context varies substantially across sessions (Figure~\ref{fig:composition}). In some sessions, tool results and file history account for over 40\% of JSONL bytes; in others, the conversation itself dominates. This explains the bimodal trim distribution: the trimmer can only remove what is there to remove.

For flat-rate subscription users, per-token costs do not apply. Trimming serves purely as a context window optimisation. While there is no direct financial penalty, it significantly extends effective session length by reducing the per-turn token footprint, which otherwise causes rate limits to deplete rapidly when repeatedly sending large, untrimmed contexts.

\subsection{Context Rebuilding: The Unquantified Cost}

The trimming analysis above captures only the marginal savings from reducing an existing session's token count. In practice, the dominant cost avoided by CMV is not trimming but \emph{context rebuilding}: the tokens, time, and output cost required to reconstruct a codebase mental model from scratch when starting a new session. The mean session in our corpus contains 84k tokens of accumulated state. Rebuilding this understanding from a blank session requires the model to re-read files, re-derive architectural relationships, and re-establish conventions---a process that in observed usage takes 10--20 user turns and 15--30 minutes of wall-clock time, with cumulative input costs growing quadratically as each turn re-sends the expanding prefix.

Branching from a snapshot eliminates this cost entirely: the model receives the full prior context in a single prompt load (\$0.53 at cache-write rates for an 84k-token session under the pricing in Table~\ref{tab:pricing}, dropping to \$0.04 on subsequent cache hits). This is the primary value proposition of the DAG model, and it is orthogonal to trimming. The evaluation in this section quantifies trimming because it is the component with a measurable trade-off (cache invalidation penalty vs.\ per-turn savings). The branching benefit---avoiding context rebuilding altogether---is harder to measure in a controlled setting but dominates the user-perceived value in practice.

\section{Limitations and Future Work}

The most significant limitation of our approach is that CMV's trimming is entirely structural rather than semantic. It removes content blindly by type (tool outputs, images, metadata) without assessing its downstream importance to the model's reasoning. If a stripped tool result is needed for subsequent reasoning, the model may hallucinate its contents or request a re-read. This is mitigated by two design choices: first, trims are applied at branch points where the new branch typically has different information needs than the source session; second, the algorithm preserves the model's own synthesis of tool outputs verbatim, so while the raw 847-line file dump is removed, the model's architectural summary of that file remains in the conversation. We have not yet quantified the impact on downstream reasoning accuracy in a controlled setting.

More broadly, the need for CMV points to a missing abstraction in current systems. AIOS \citep{mei2025aios} proposes an LLM Agent Operating System that embeds language models into the OS layer, with kernel-level services for scheduling, context management, memory, and access control. Their architecture treats LLM instances as cores (analogous to CPU cores) and agent requests as system calls. This is the right direction, but the specific problem of persistent, version-controlled conversational state across sessions remains underspecified in their framework. AIOS provides context snapshot and restoration within a single agent lifecycle; it does not address named branching points, DAG-based lineage, or conversationally lossless trimming of accumulated session state. CMV provides empirical evidence for this gap. The fact that context reuse, branching, and trimming had to be built in userland on top of JSONL files and filesystem copies motivates the inclusion of persistent conversational state management as a first-class concern in future agent OS designs.

The evaluation is a single-user case study; results may not generalise across usage patterns, codebases, or programming styles. The byte-to-token estimation used in the benchmark overestimates reduction for image-heavy sessions due to the discrepancy between base64 encoding size and API vision-token cost. Future work includes: (1) controlled comparisons of trimmed vs.\ untrimmed branches given identical follow-up tasks to quantify any downstream reasoning degradation; (2) multi-user evaluation across diverse usage patterns; (3) adaptive trim thresholds informed by auto-trim log data; and (4) exploration of how CMV's DAG state model and trimming algorithm might serve as a reference implementation for the persistent context subsystem in an AIOS-style architecture.
\section{Conclusion}

Contextual memory virtualisation provides a principled framework for treating LLM conversational state as a persistent, version-controlled resource rather than ephemeral session data. The DAG-based state model enables context reuse patterns (branching, chaining, team sharing) that are impossible under the current session-per-task paradigm. The three-pass trimming algorithm achieves significant token reduction while maintaining both conversational completeness and API correctness. Economic analysis confirms viability. API users recover the prompt caching penalty within a small number of turns, and subscription users gain pure context window savings. Ultimately, as agents are deployed for increasingly complex, multi-day tasks, routinely discarding hours of accumulated context due to window limits becomes an unacceptable overhead. CMV shows that we do not need to wait for model-level breakthroughs or endlessly expanding context windows to fix this; we can solve context ephemerality right now at the tooling layer.

\bibliographystyle{plainnat}

\end{document}